\documentclass[epjCONF]{svjour}
\usepackage{graphics}
\usepackage[varg]{txfonts} % Times fonts
\usepackage[latin1]{inputenc}
\session-title{Meson2012 - 12th International Workshop on Production, Properties and Interaction of Mesons}
\begin{document}
\title{In-medium Properties of Hadrons}
\author{Volker Metag\inst{1} \fnmsep\thanks{\email{Volker.Metag@exp2.physik.uni-giessen.de}}\\ }
\institute{II. Physikalisches Institut, University of Giessen, Germany}
\abstract{
An overview is given over recent results on in-medium properties of hadrons, obtained in experiments with photon, proton and deuteron beams at ANKE, CBELSA/TAPS, Crystal Ball, HADES, and BigRIPS. These experiments focus on identifying spectral in-medium modifications of hadrons, frequently discussed in the context of a partial restoration of chiral symmetry at finite nuclear densities. Three experimental approaches are presented: the measurement of the transparency ratio, the meson line shape analysis, and the search for meson-nucleus bound states. Results for $\omega,\phi, $ and $ \eta^\prime$ mesons indicate a broadening in the nuclear medium. Corresponding inelastic in-medium meson-nucleon cross sections have been extracted.  Evidence for an in-medium mass shift has not been reported. Further information on the meson-nucleus interaction is derived from a spectroscopy of meson-nucleus bound states. A critical comparison of the results with theoretical predictions is presented.\ \
} %end of abstract
\maketitle
\section{Introduction}
\label{intro}
The origin of hadron masses and predicted in-medium modifications of hadron properties are key issues in Quantum Chromodynamics (QCD) in the strong coupling regime.  Widespread theoretical and experimental studies focus on the question whether well known hadron properties change when these hadrons are embedded in a strongly interacting environment. Spectral modifications of hadrons, encoded in changes of their mass and decay width, are often discussed in the context of a restoration of the broken chiral symmetry in nuclei. However, as shown in \cite{Leupold:2009kz}, the connection between QCD symmetries and hadron properties are much more involved than initially thought. As a consequence, hadronic models, based on our current understanding of meson-baryon interactions but constrained by symmetry considerations, have been used by several theory groups to calculate the in-medium self-energies of hadrons and their spectral functions. 

\begin{figure}[tb]
\begin{center}
%\resizebox{0.28\columnwidth}{!}{\includegraphics{hatsuda_last.eps}}
\resizebox{0.25\columnwidth}{!}{\includegraphics{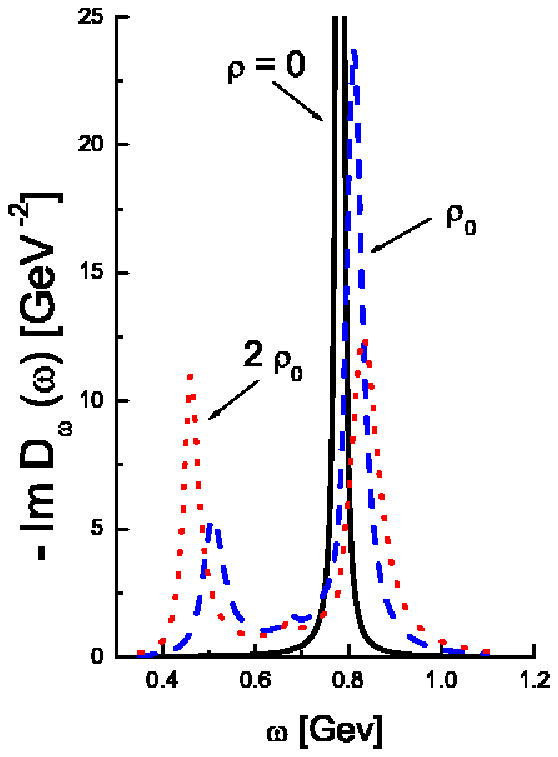}}
\resizebox{0.42\columnwidth}{!}{\includegraphics{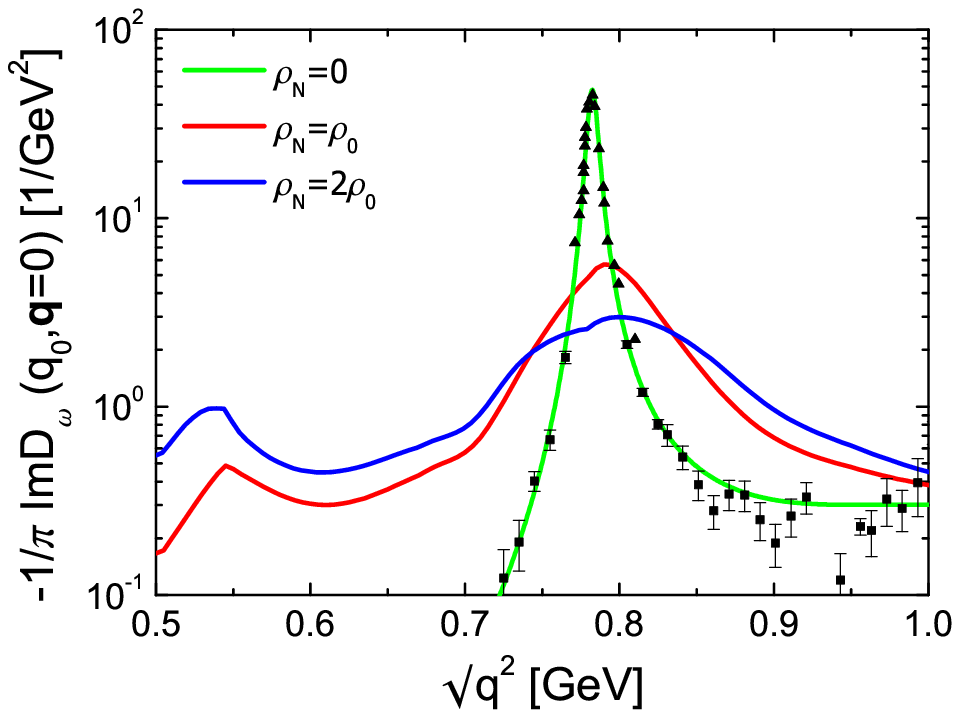}}
\resizebox{0.22\columnwidth}{!}{\includegraphics{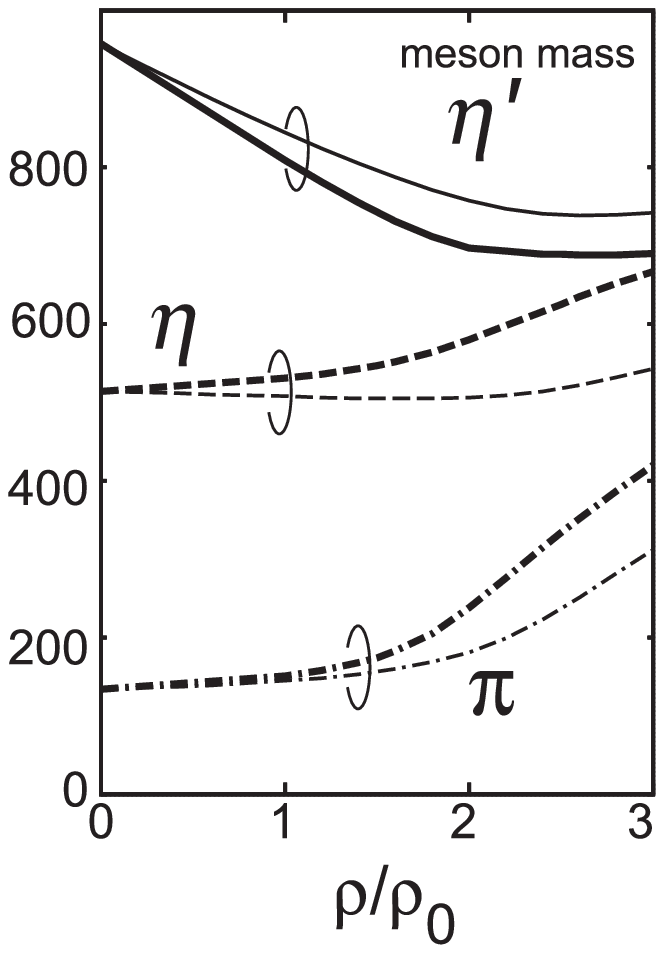}}
\caption{Theoretical predictions for in-medium modifications of mesons: (Left) Structures in the $\omega$ spectral function arising from the coupling of the $\omega$ meson to nucleon resonances \cite{Lutz}. (Middle) In-medium broadening of the $\omega$ meson calculated in a coupled channel approach \cite{Muehlich_omega_sf}. (Right) Density dependence of pseudoscalar meson masses calculated for SU(2) symmetric matter (thick lines) and SU(3) symmetric matter (thin lines) \cite{Nagahiro}.}
\label{fig:theo}
\end{center}
\end{figure}

Fig.~\ref{fig:theo} shows examples for different predictions of in-medium modification scenarios. Lutz et al. \cite{Lutz} calculate a slight upward shift in mass and a structure in the $\omega$ meson spectral function near 500 MeV/c$^2$, arising from the coupling of the $\omega$ meson to nucleon resonances. M\"uhlich et al. \cite{Muehlich_omega_sf} predict a strong broadening of the $\omega$ meson at normal and twice normal nuclear-matter density. Hatsuda and Lee \cite{Hatsuda_Lee}, applying a QCD sum-rule approach, predict a lowering of vector meson masses by 10-20$\%$ at normal nuclear-matter density for $\rho$ and $\omega$ mesons while for the $\phi$ meson only a small mass drop is expected due to the only weak interaction of $\phi$ mesons with nuclei (not shown in Fig.~\ref{fig:theo}). Within the NJL model, Nagahiro et al. \cite{Nagahiro} calculate a lowering of the $\eta^\prime$ mass by about 15$\%$ at normal nuclear matter density, in contrast to Bernard and Meissner \cite{meissner} who expect a rather weak density dependence of the $\eta^\prime$ mass (not shown in Fig.~\ref{fig:theo}). An experimental test of these theoretical predictions thus requires measurements sensitive to mass shifts, structures and/or broadening of hadronic spectral functions and detectors with acceptance for low momentum mesons. A direct comparison of theoretical predictions and experimental results is, however, hampered by the fact that most of these calculations have been performed for idealized conditions, assuming mesons at rest in infinitely extended nuclear matter in equilibrium at constant density and temperature - a scenario very different from the production of mesons in a nuclear reaction on finite nuclei. To allow for a quantitative comparison, predicted in-medium modifications have to be implemented as input into transport calculation, e.g. GIBUU calculations \cite{Buss}, which take effects into account like:
\begin{itemize}
\item initial state effects: absorption of incoming beam particles
\item non-equilibrium effects: varying density and temperature
\item absorption and regeneration of mesons
\item fraction of meson decays outside of the nuclear environment
\item final state interactions: distortion of 4-momenta of meson decay products
\end{itemize}
A comparison of theoretical predictions and experimental observables thus always requires the use of transport calculations. In the following sections three experimental approaches are described which have been applied in experiments at different accelerator facilities to extract in-medium properties of mesons. 

\section{Experimental approaches}
\label{sec:1}
\subsection{Measurement of the transparency ratio}
\label{sec:2}
\begin{figure}[tb]
\begin{center}
 \resizebox{0.60\columnwidth}{!}{\includegraphics{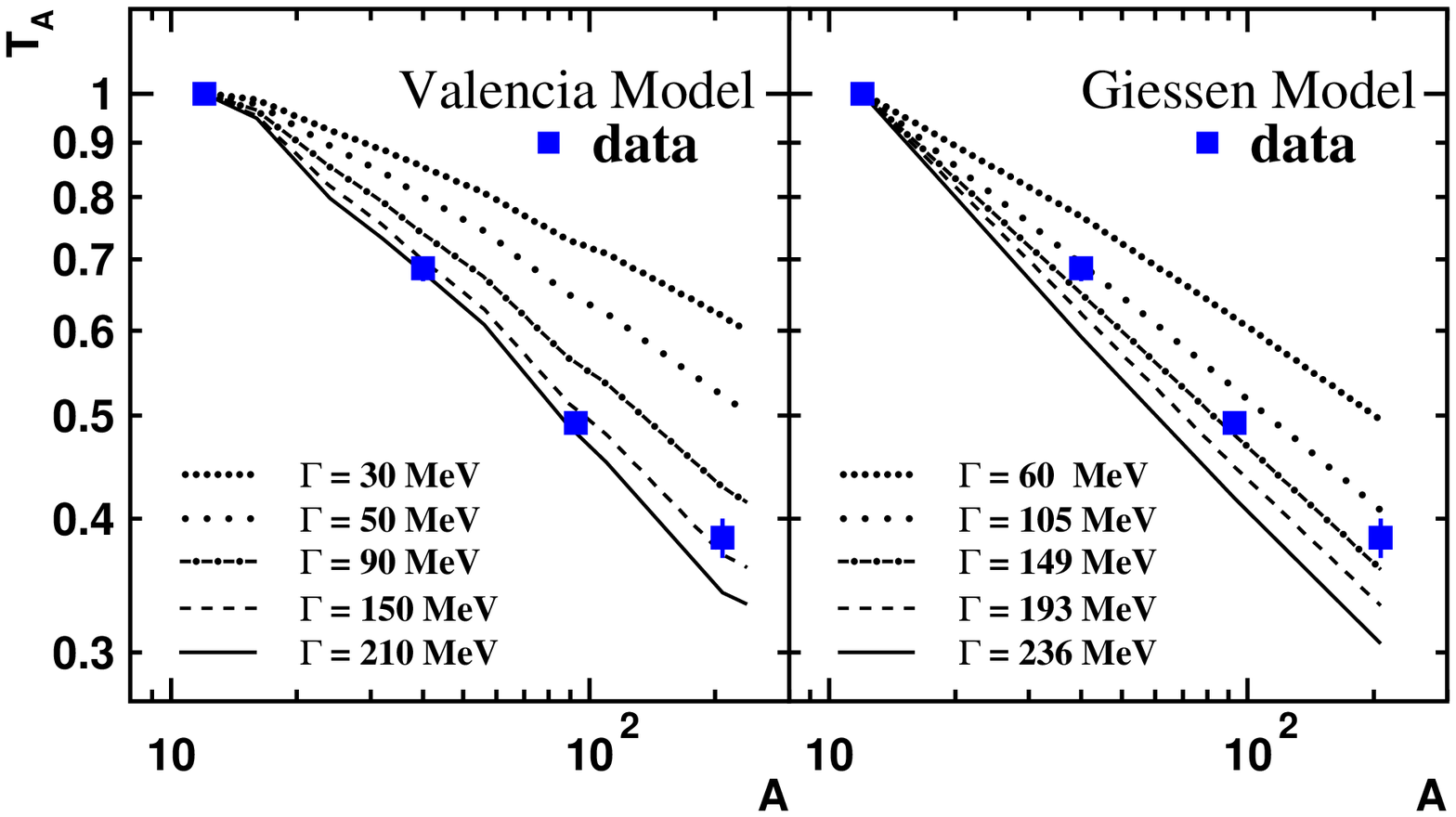}}
  \resizebox{0.38\columnwidth}{!}{\includegraphics{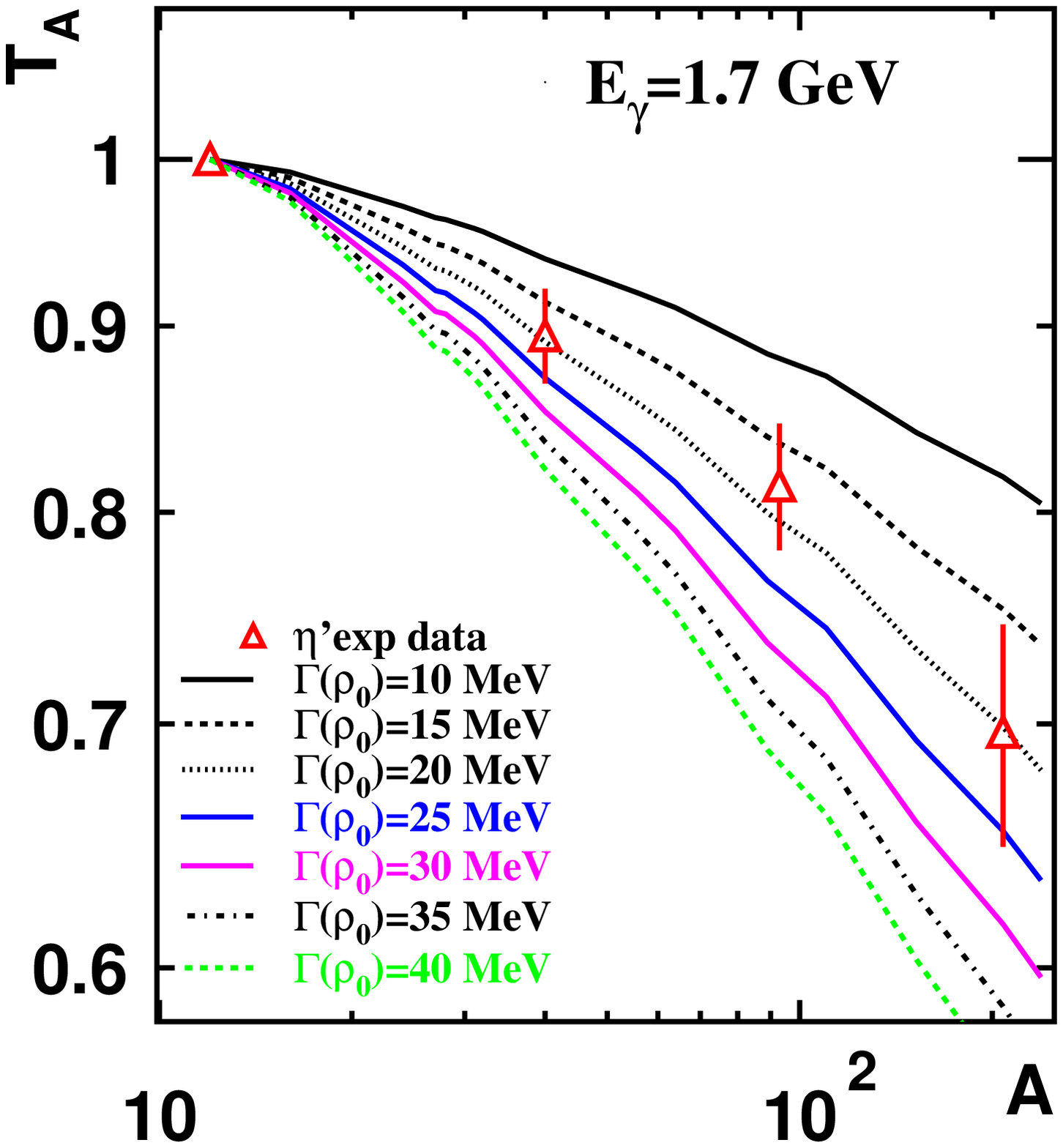}}
\caption{Transparency ratios for different nuclei, normalized to C, measured in photo nuclear reactions for the $\omega$ meson (left) \cite{Kotulla} and the $\eta^\prime$ meson (right) \cite{Mariana_etaprime} in comparison to transport calculations \cite{Muehlich_transp,Kaskulov_transp} for different in-medium widths.
\label{fig:transp}}
\end{center}
\end{figure}

The transparency ratio, defined here for the $\eta^\prime$ meson as
\begin{equation}
T_A=\frac{ \sigma_{\gamma A\to \eta^\prime A^\prime}}{{A}\cdot\sigma_{\gamma N\to \eta^\prime X} } \ .\label{eq:trans}
\end{equation} 
 is a measure for the absorption of mesons in a nucleus: the production cross section per nucleon within a nucleus is compared to the meson production cross section on a free nucleon. As discussed in \cite{Mariana_etaprime} the transparency ratio is, however, usually normalized to a light nucleus like carbon. The removal of mesons by inelastic processes leads to a shortening of the meson lifetime within the nucleus and thus to an increase in width. In the low density approximation, the meson width in the medium and the inelastic meson-nucleon cross section are related by 
 \begin{equation}
\Gamma = \hbar c \cdot \rho_{0} \cdot \sigma_{\rm inel} \cdot \beta, \label{eq:Gamma-sigma} 
%\end{equation}
\hspace*{1cm} with  \hspace*{0.5cm} 
%\begin{equation}
\beta = \frac{p_{\eta^\prime}}{E_{\eta^\prime}}
\end{equation}
Fig.~\ref{fig:transp} shows transparency ratios measured in photo nuclear reactions by the CBELSA/TAPS collaboration for $\omega$ \cite{Kotulla} and $\eta^\prime$ mesons \cite{Mariana_etaprime}. A comparison to transport calculations \cite{Mariana_etaprime,Mariana_proc,Muehlich_transp} gives widths of 130 -150 MeV and 15-25 MeV, respectively, corresponding to inelastic meson-nucleon cross sections of $\approx$ 60 mb and 3-10 mb.
\begin{figure}[tb]
\begin{center}
 \resizebox{0.27\columnwidth}{!}{ \includegraphics{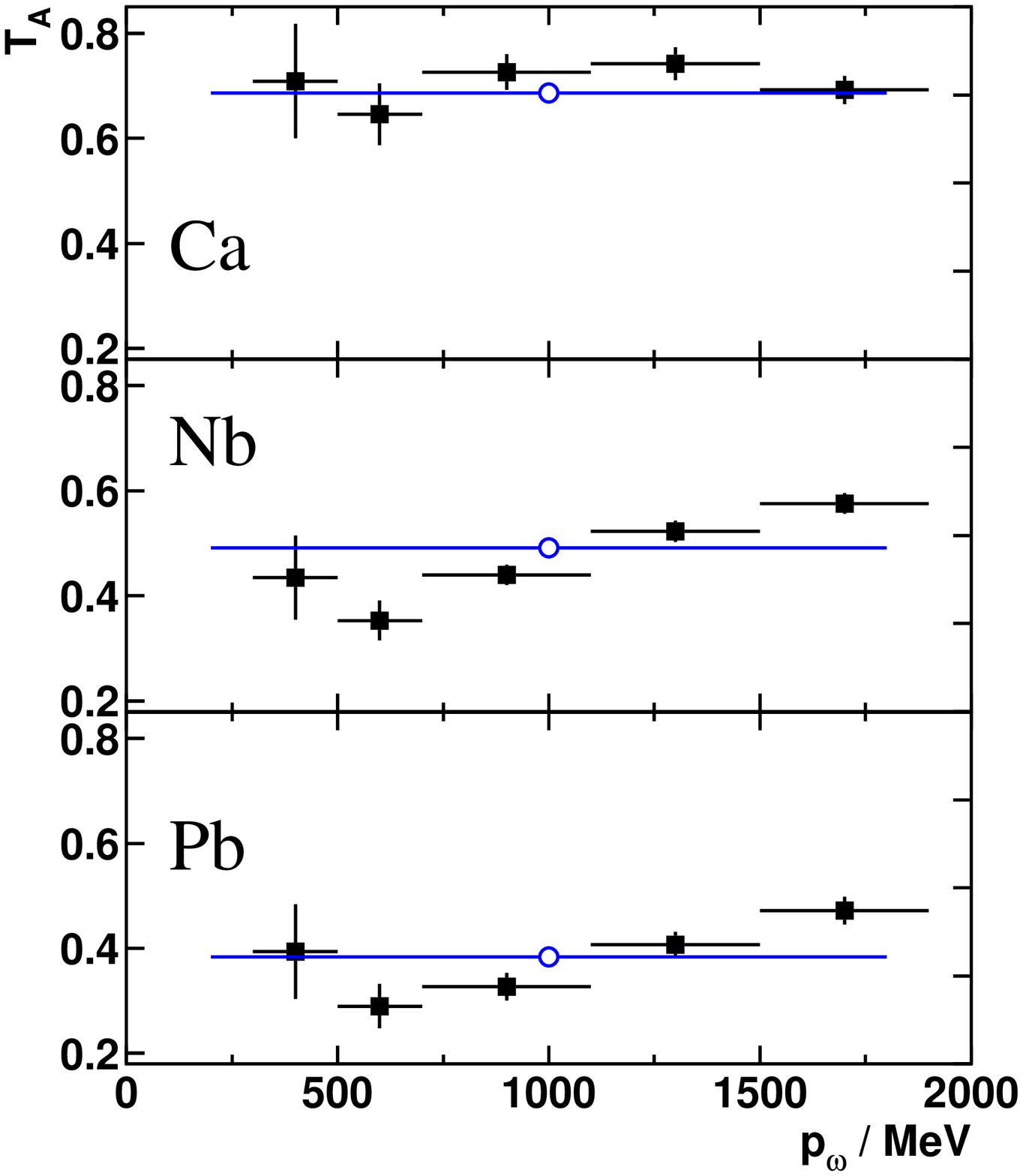}}3b
 \resizebox{0.35\columnwidth}{!}{\includegraphics{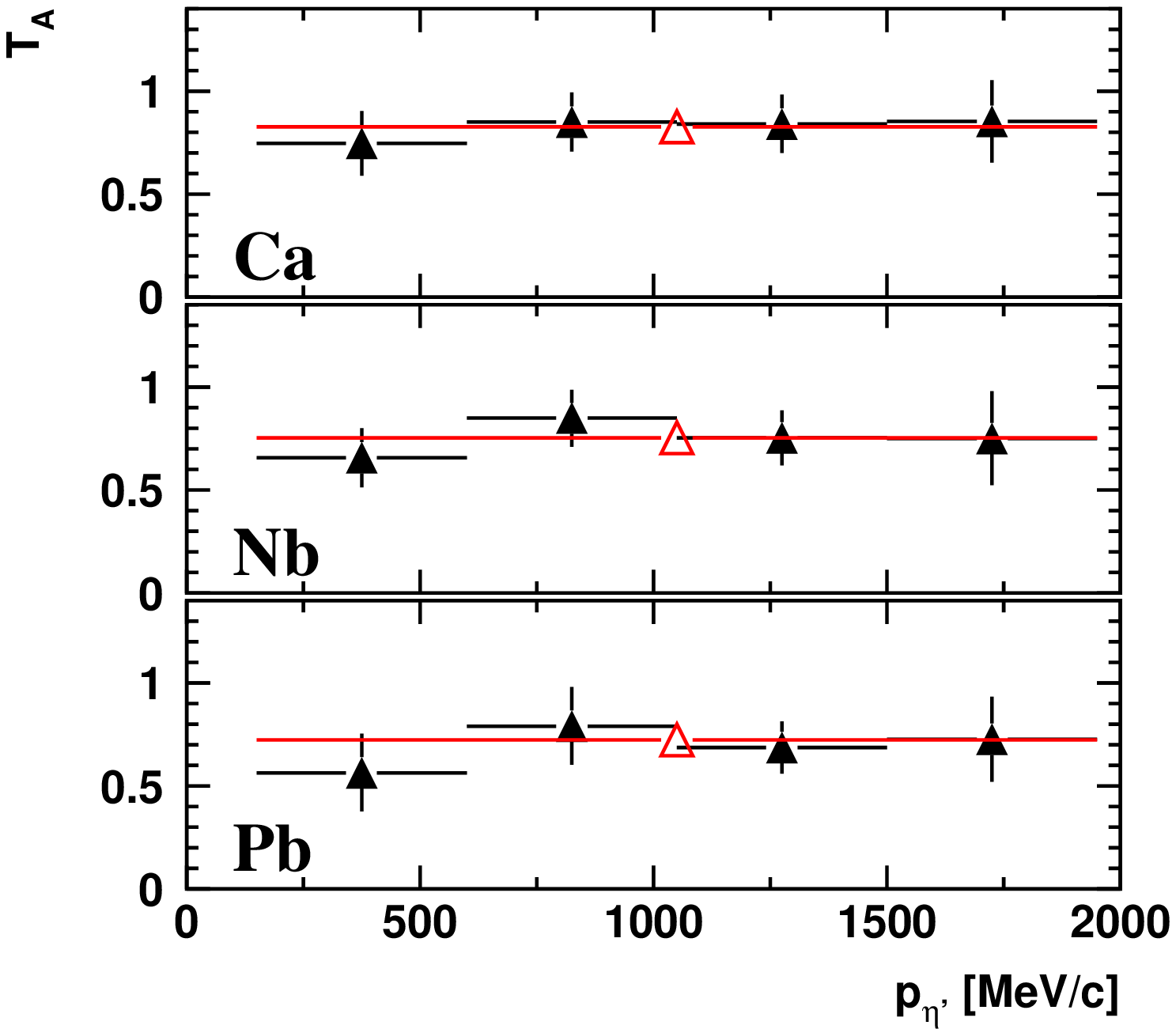}}
 \resizebox{0.32\columnwidth}{!}{\includegraphics{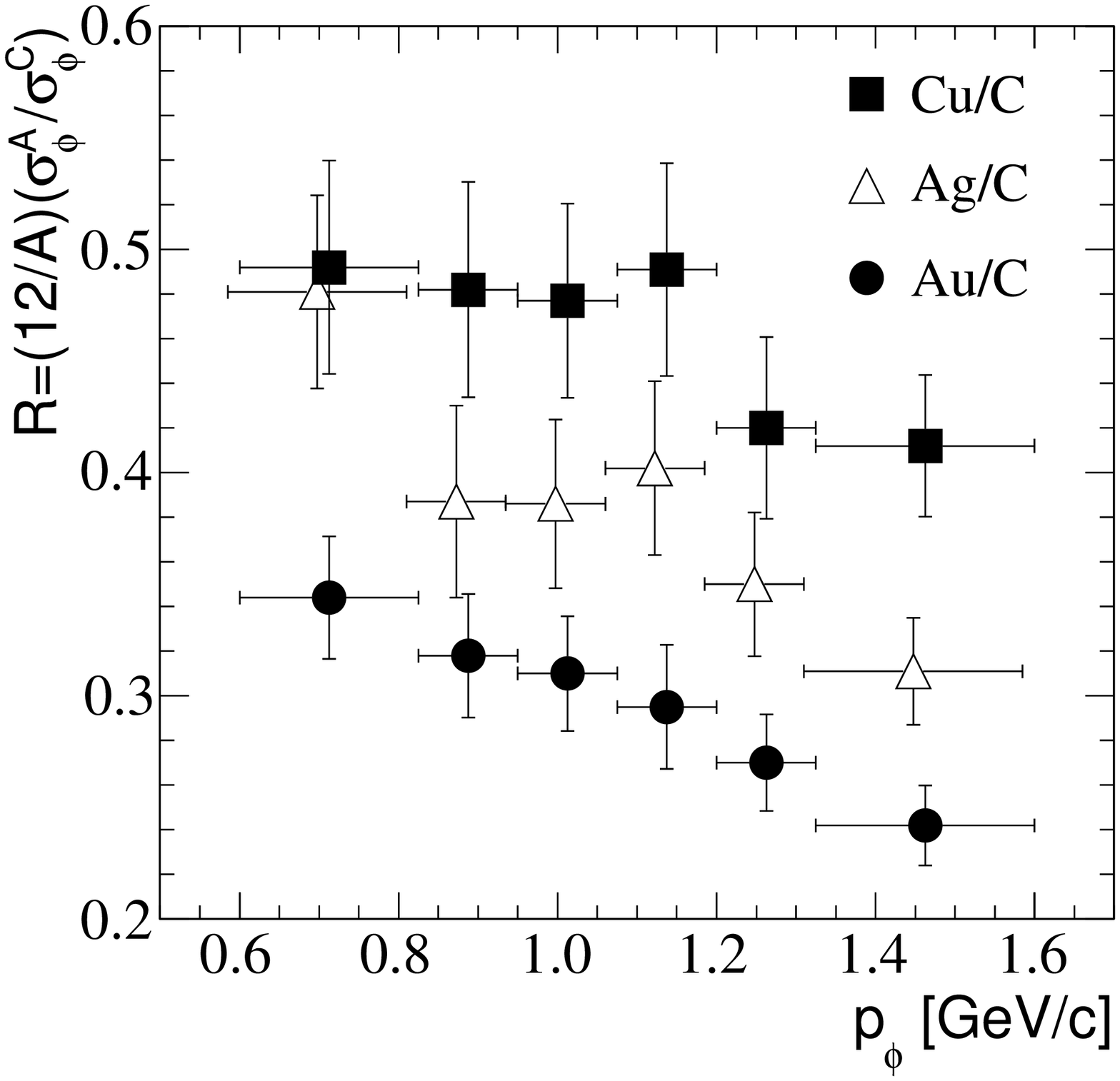}}
\caption{Momentum dependence of the transparency ratio for $\omega$ (left) \cite{Kotulla}, $\eta^\prime$ \cite{Mariana_etaprime}, and $\phi$ mesons \cite{Hartmann,Polyanskiy}.
\label{fig:transp_mom}}
\end{center}
\end{figure}
\begin{figure}[tb]
\begin{center}
\resizebox{0.33\columnwidth}{!}{\includegraphics{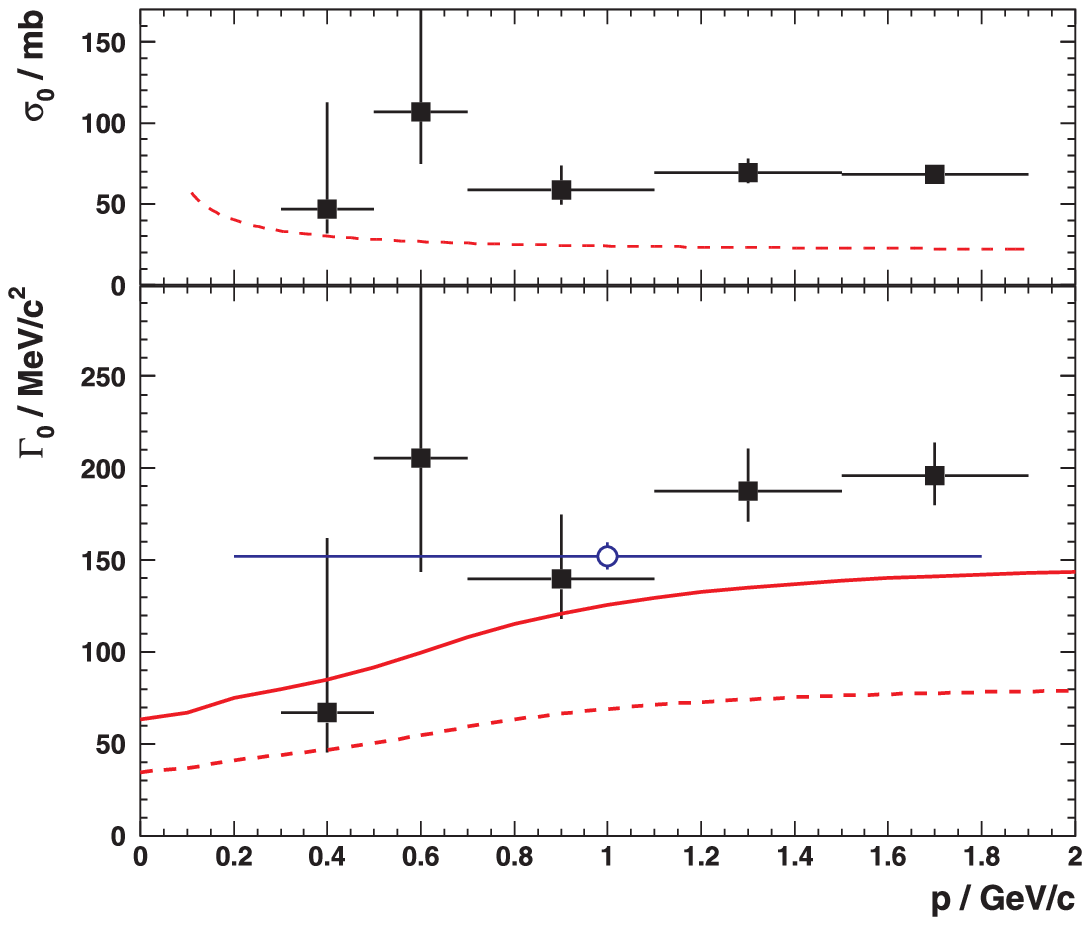}}
 \resizebox{0.40\columnwidth}{!}{\includegraphics{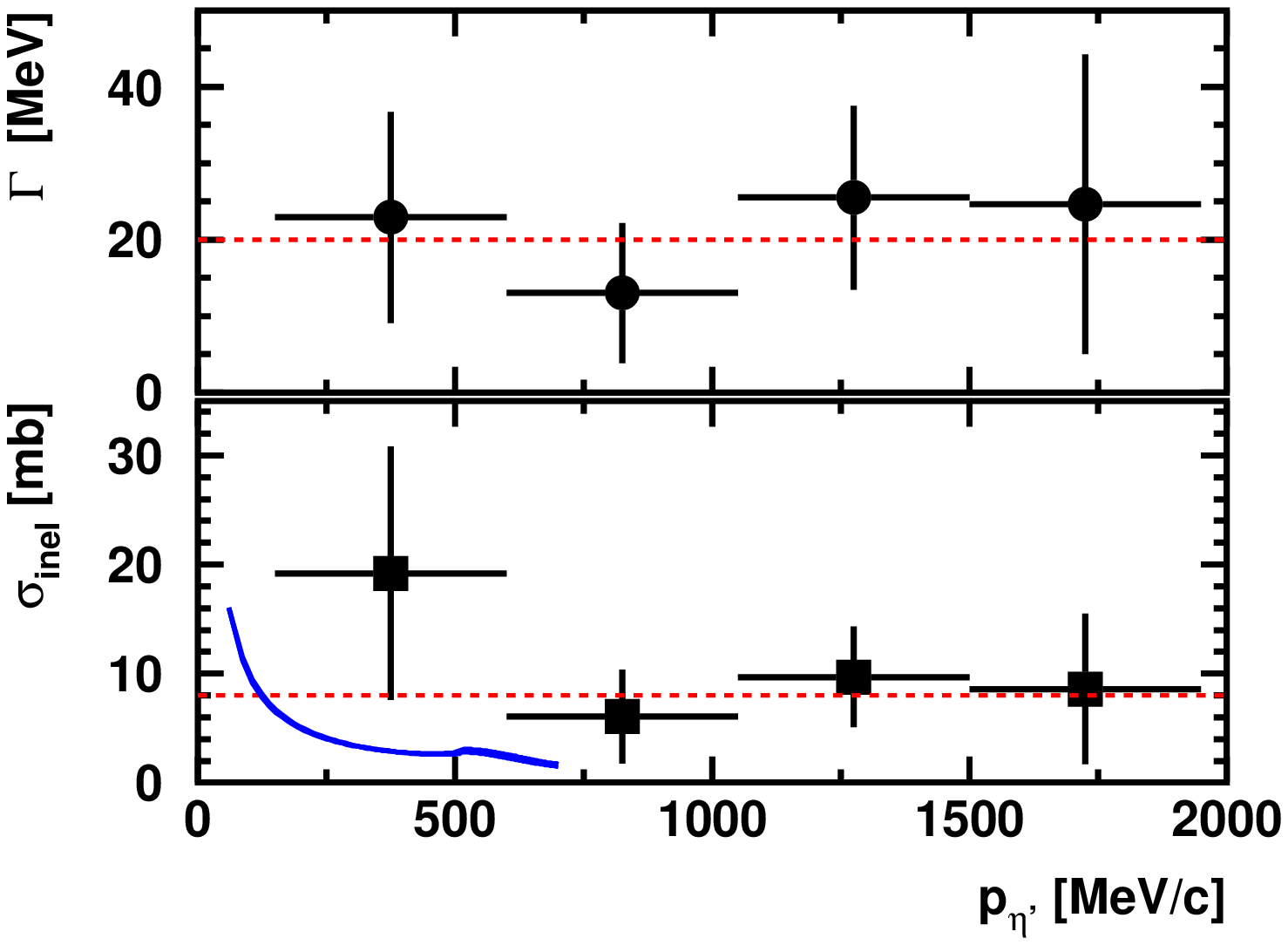}}
 \resizebox{0.25\columnwidth}{!}{\includegraphics{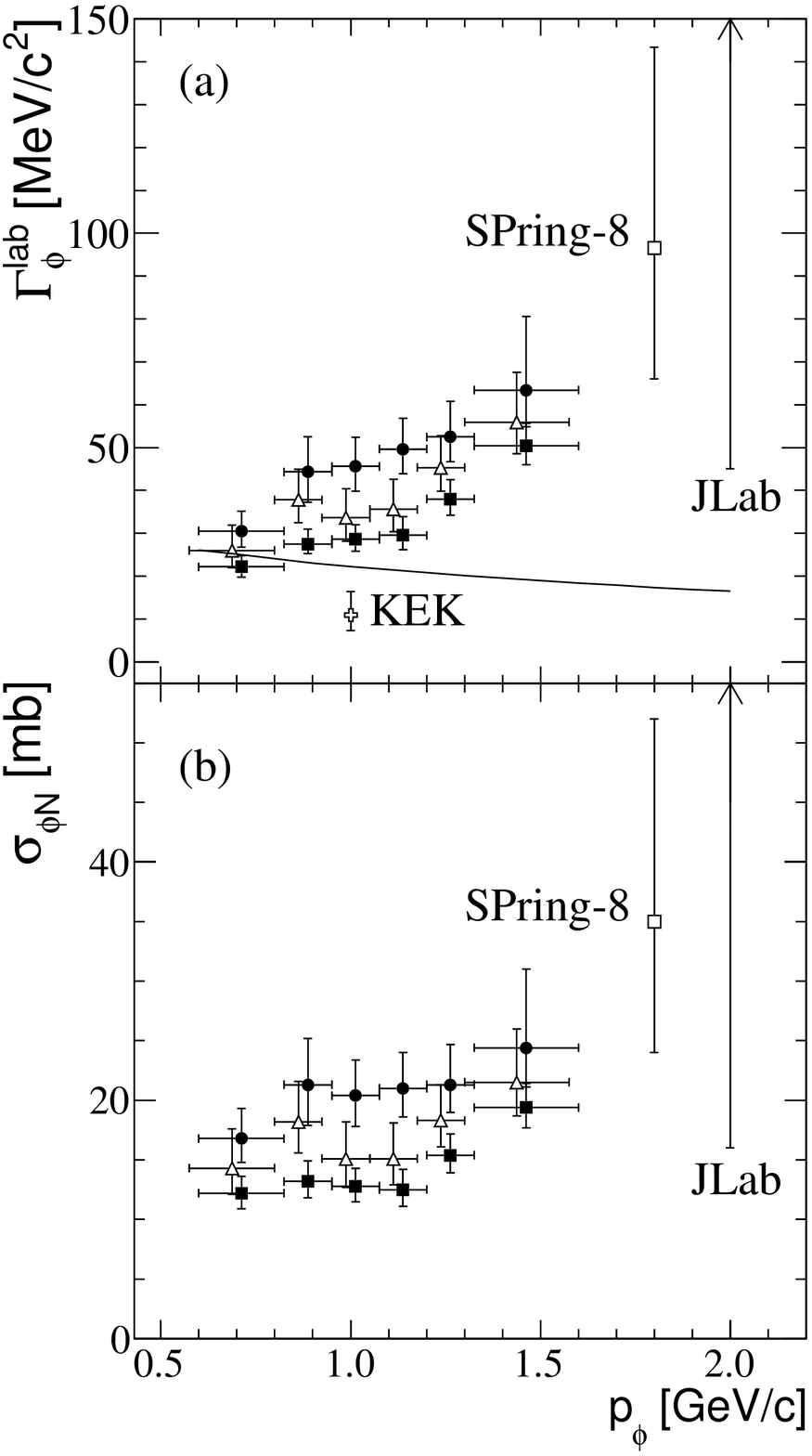}}
\caption{In-medium widths and inelastic meson-nucleon cross sections for $\omega$ (left) \cite{Kotulla}, $\eta^\prime$ \cite{Mariana_etaprime,Mariana_proc}, and $\phi$ mesons \cite{Hartmann,Polyanskiy}.
\label{fig:Gamma_cs}}
\end{center}
\end{figure}
Two-step-processes in meson production, like e.g., $\gamma N \rightarrow N \pi$ followed by $\pi N \rightarrow \eta^\prime N$ will distort the absorption measurement and can even lead to transparency ratios larger than 1. Since mesons produced in secondary reactions are predominantly low energetic, a momentum dependence of the transparency ratio is indicative for such processes. Results on the momentum dependence for $\omega$ \cite{Kotulla}, $ \eta^\prime$ \cite{Mariana_etaprime,Mariana_proc} and $\phi$ mesons \cite{Hartmann,Polyanskiy} are shown in Fig.~\ref{fig:transp_mom}. While the transparency ratio is almost constant for $\omega$ and $\eta^\prime$ mesons there is a momentum dependence  in case of the $\phi$ meson, indicating contributions from two-step-processes. The corresponding in-medium widths and related inelastic cross sections are plotted as a function of the meson momentum in  Fig.~\ref{fig:Gamma_cs}.

\begin{figure}[tb]
\begin{center}
\resizebox{0.37\columnwidth}{!}{\includegraphics{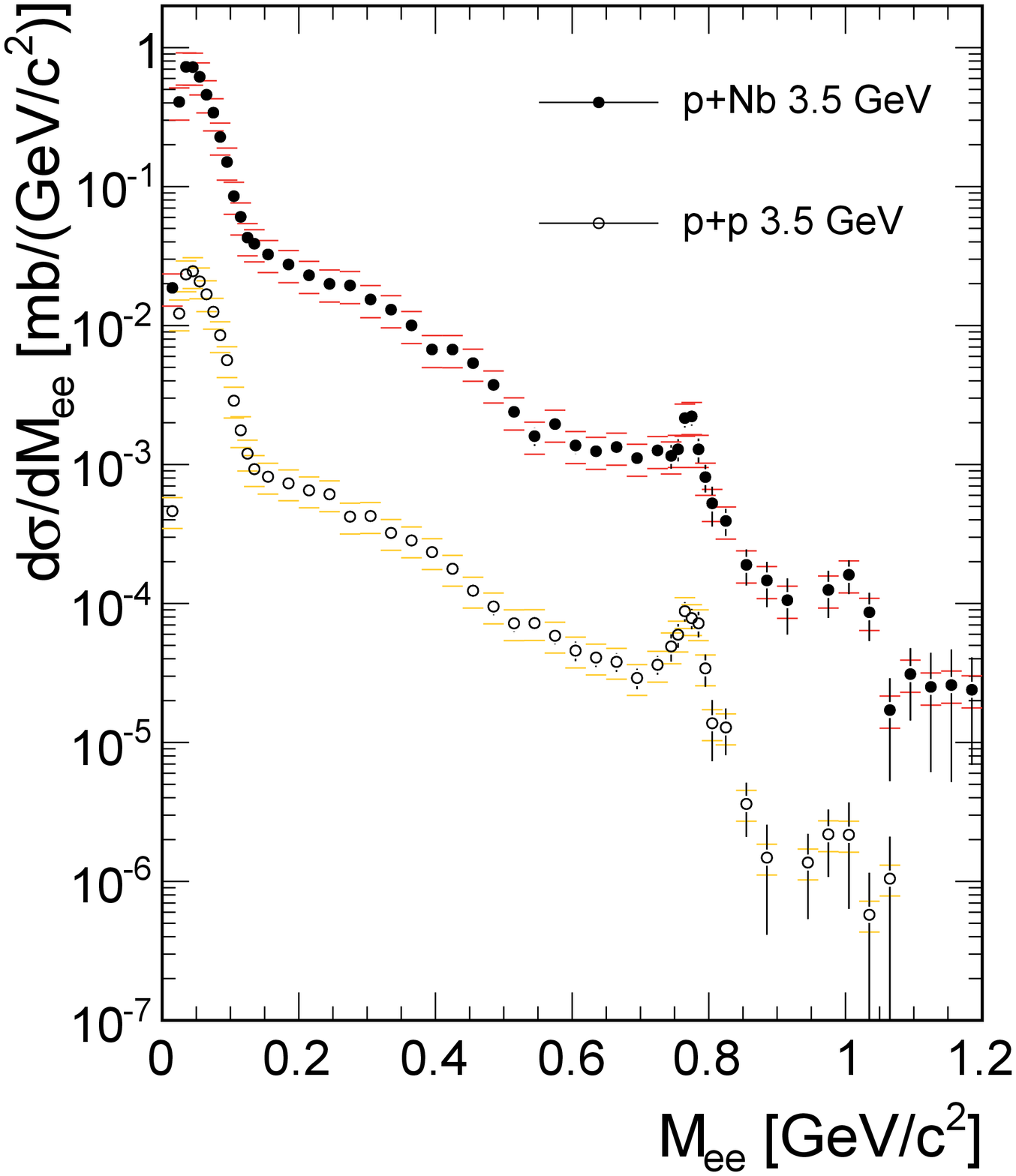}}
\resizebox{0.41\columnwidth}{!}{\includegraphics{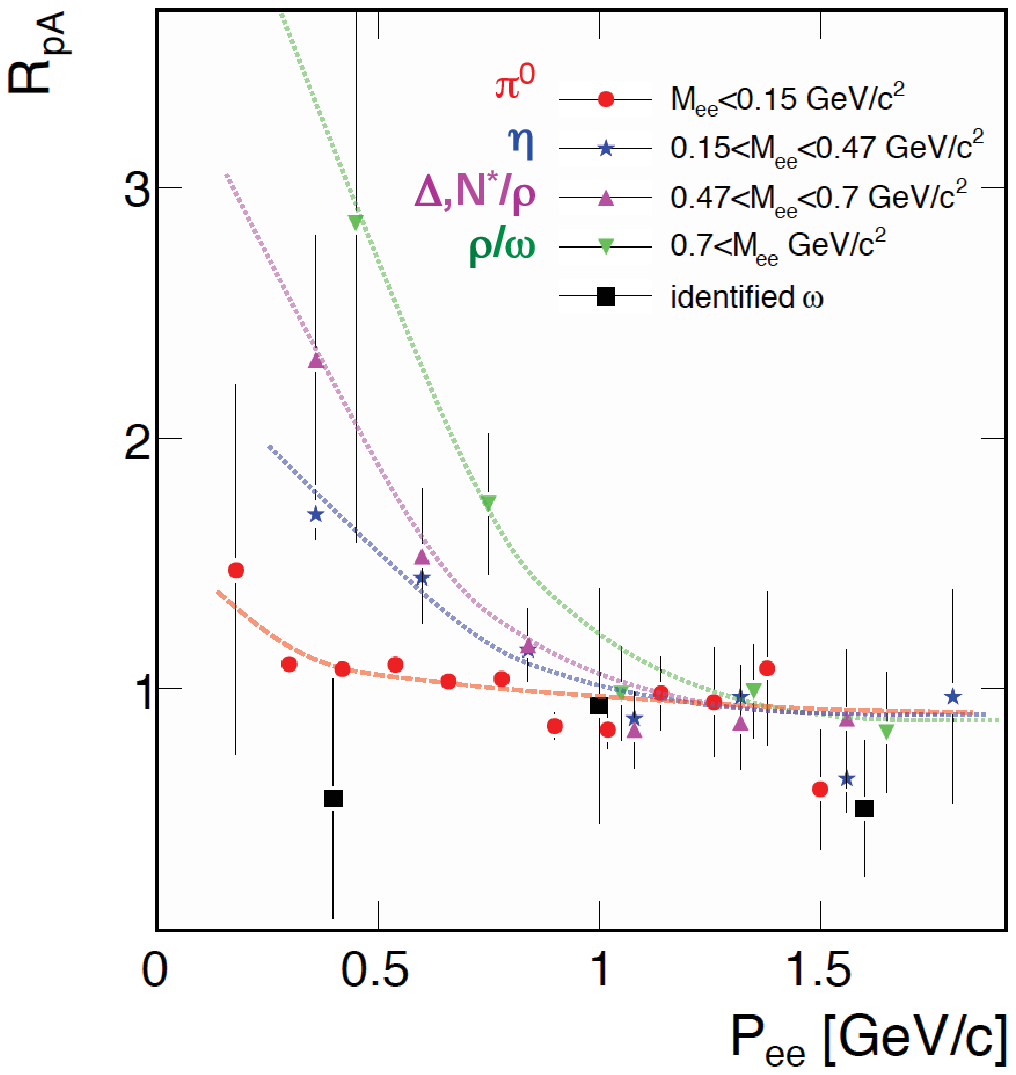}}
\caption{(Left) Comparison of dielectron cross sections as a function of the invariant mass in p+p and p+Nb collisions at 3.5 AGeV. (Right) The transparency ratio R$_{pA}$ as a function of the $e^+e^-$ pair momentum for different $e^+e^-$ mass bins \cite{Lorenz} \label{fig:e+e-_mass_RpA}.}
\end{center}
\end{figure}

The HADES collaboration has measured $e^+e^-$ pairs from p+p and p+Nb reactions (see Fig.~\ref{fig:e+e-_mass_RpA} (left)) \cite{Lorenz}. The meson transparency ratio is here expressed by the equivalent quantity R$_{pA}$ (Fig.~\ref{fig:e+e-_mass_RpA} (right)). Values larger than 1 indicate contributions from two-step production processes which become more and more important the larger the $e^+e^-$ mass. For the $\omega$ meson a suppression consistent with the CBELSA/TAPS result \cite{Kotulla} is observed.

\subsection{Analysis of the meson line shape}
\label{sec:3}
\begin{figure}[tb]
\begin{center}
\resizebox{0.36\columnwidth}{!}{\includegraphics{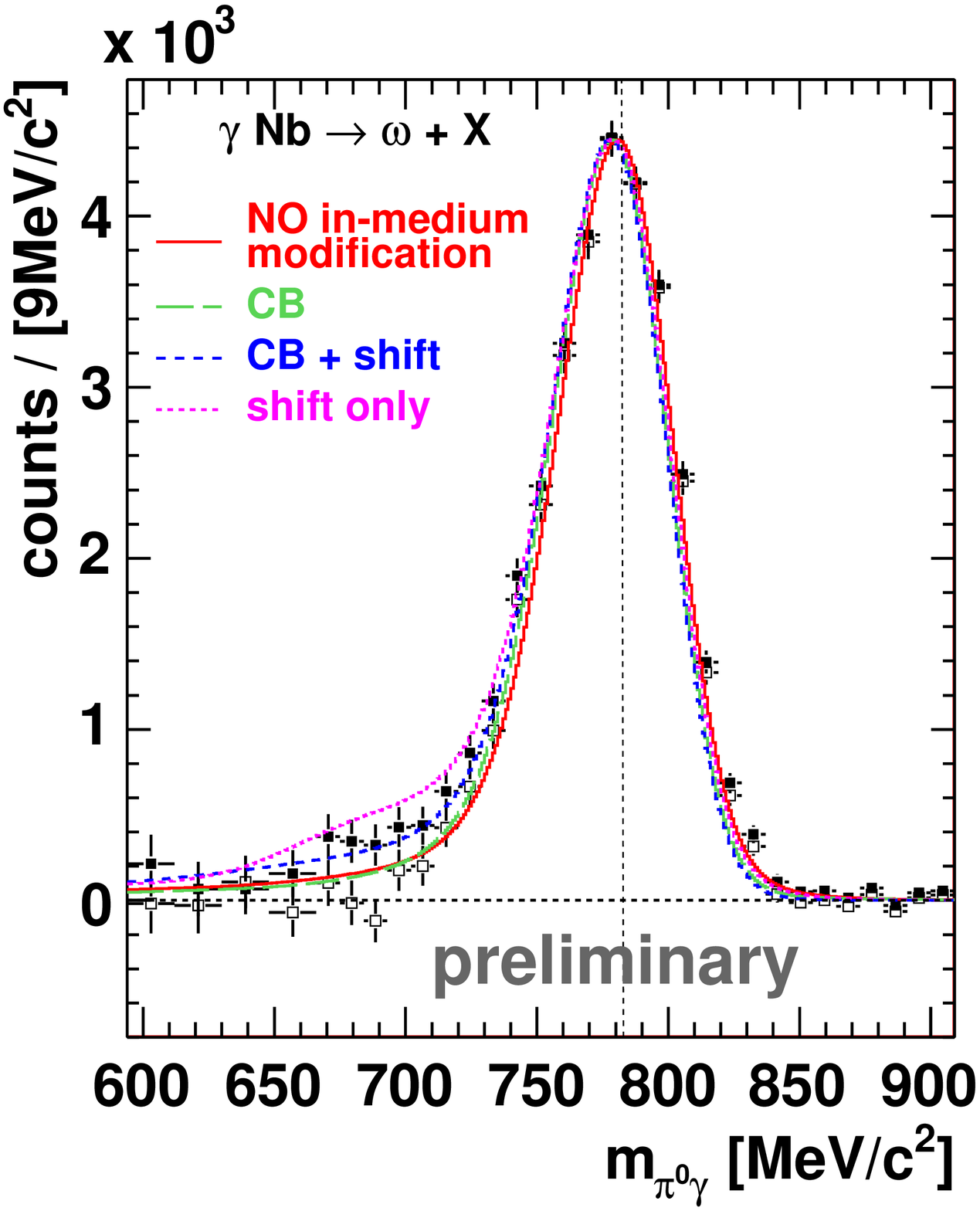}}
 \resizebox{0.39\columnwidth}{!}{\includegraphics{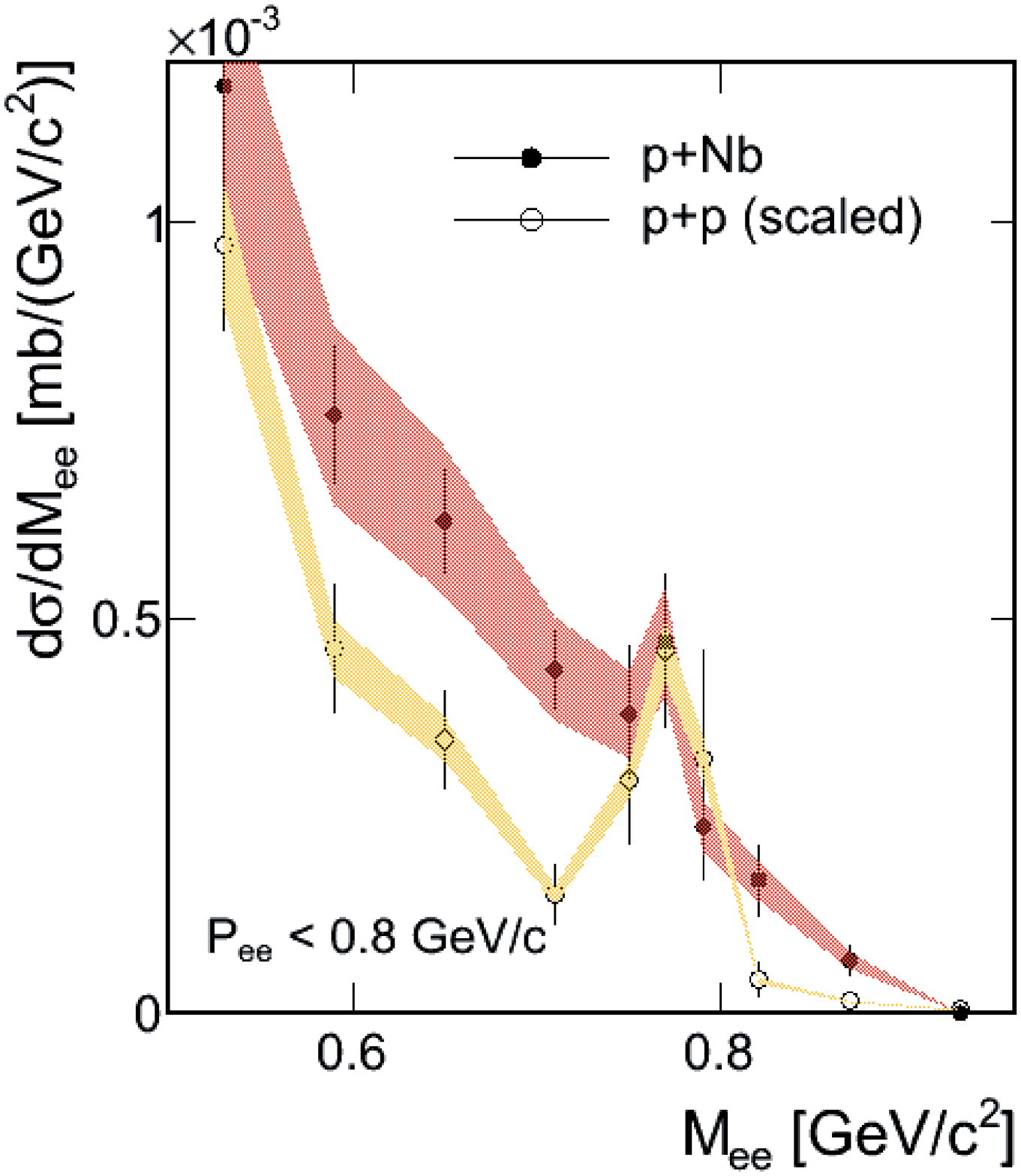}}
\caption{$\omega$ line shape measured in the $\omega \rightarrow \pi^0 \gamma$ (left) \cite{Thiel} and $\omega \rightarrow e^+e^-$ decay channels \cite{Lorenz}. The two sets of data points (left) have been obtained by applying different background subtraction methods and are compared to GiBUU calculations for different assumed in-medium modification scenarios \cite{Mariana_EPJA,Weil}. \label{fig:lineshape}}
\end{center}
\end{figure}

The line shape of the $\omega$ meson has been studied in the decay modes $\omega \rightarrow \pi^0 \gamma $ \cite{Mariana_PRC,Mariana_EPJA,Thiel} and 
$\omega \rightarrow e^+e^-$ \cite{Lorenz}. The preliminary $\omega \rightarrow \pi^0 \gamma $ signal obtained in recent photo production experiments (E$_{\gamma} = 900- 1300 $~MeV) at MAMI-C on a Nb target \cite{Thiel} is shown in  Fig.~\ref{fig:lineshape} (left). The two sets of data points reflect systematic uncertainties in the background subtraction \cite{Metag_Erice_11}. These uncertainties are comparable in magnitude to the predicted differences in the $\omega$ line shape for various in-medium modification scenarios \cite{Mariana_EPJA,Weil} and thus do not allow to distinguish between them; only the case of a mass shift without broadening appears not to be consistent with the data.  Fig.~\ref{fig:lineshape} (right) shows the $e^+e^-$ invariant mass spectrum measured by HADES in the vector meson mass range \cite{Lorenz}. Comparing the p+Nb data to a p+p reference measurement, an excess $e^+e^-$ yield is observed which is attributed to additional contributions from nucleon resonances decaying via $N^* \rightarrow N \rho \rightarrow N e^+e^-$. The extraction of the width of the $\omega$ peak is hampered by low statistics due to the strong $\omega$ absorption in case of the p+Nb data. If any in-medium broadening occurs, the change in width of the observed signal is on the percentage level.

\subsection{Search for meson-nucleus bound states}
\begin{figure}[h!]
 \begin{center}
\resizebox{0.48\columnwidth}{!}{\includegraphics{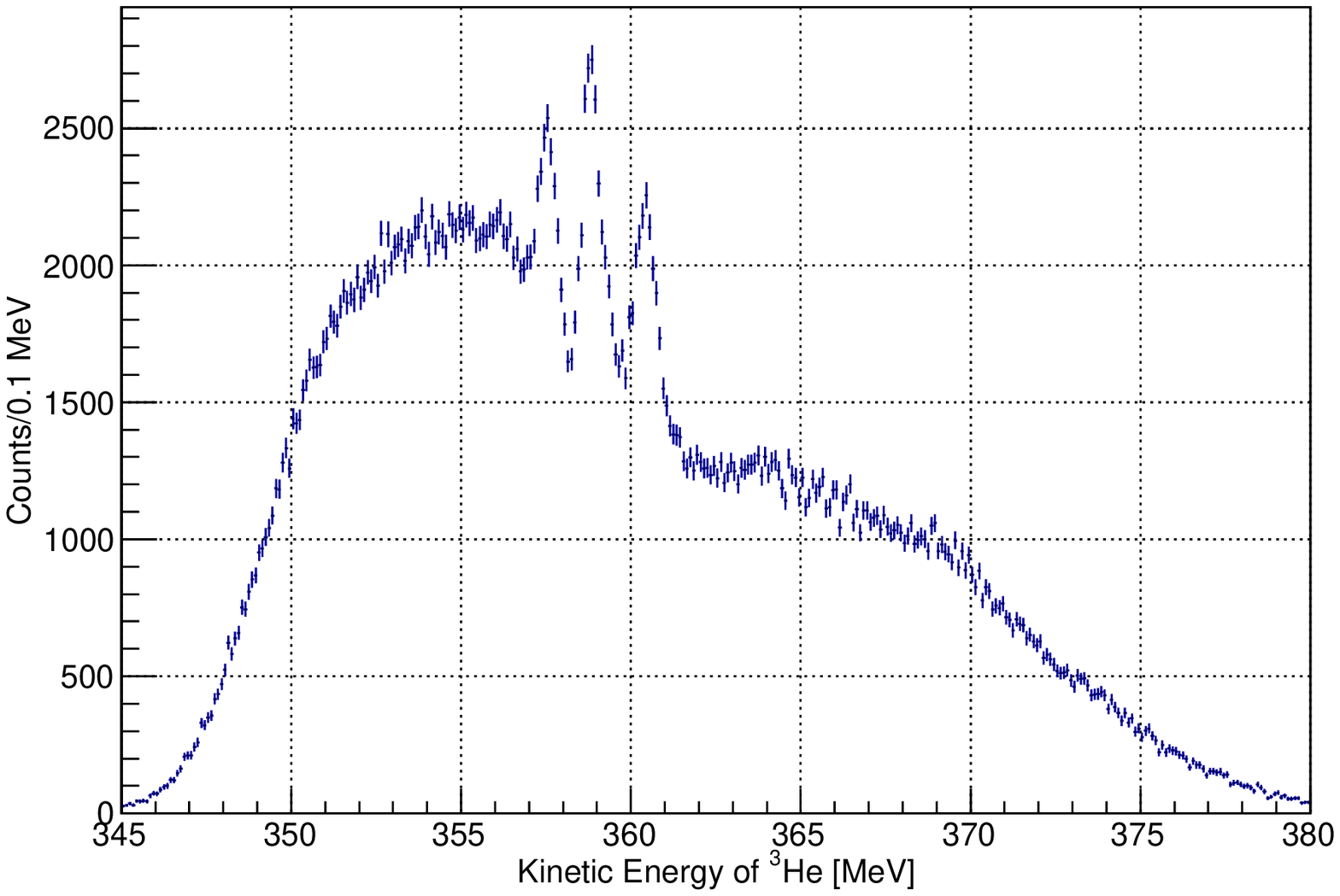}}
 \resizebox{0.46\columnwidth}{!}{\includegraphics{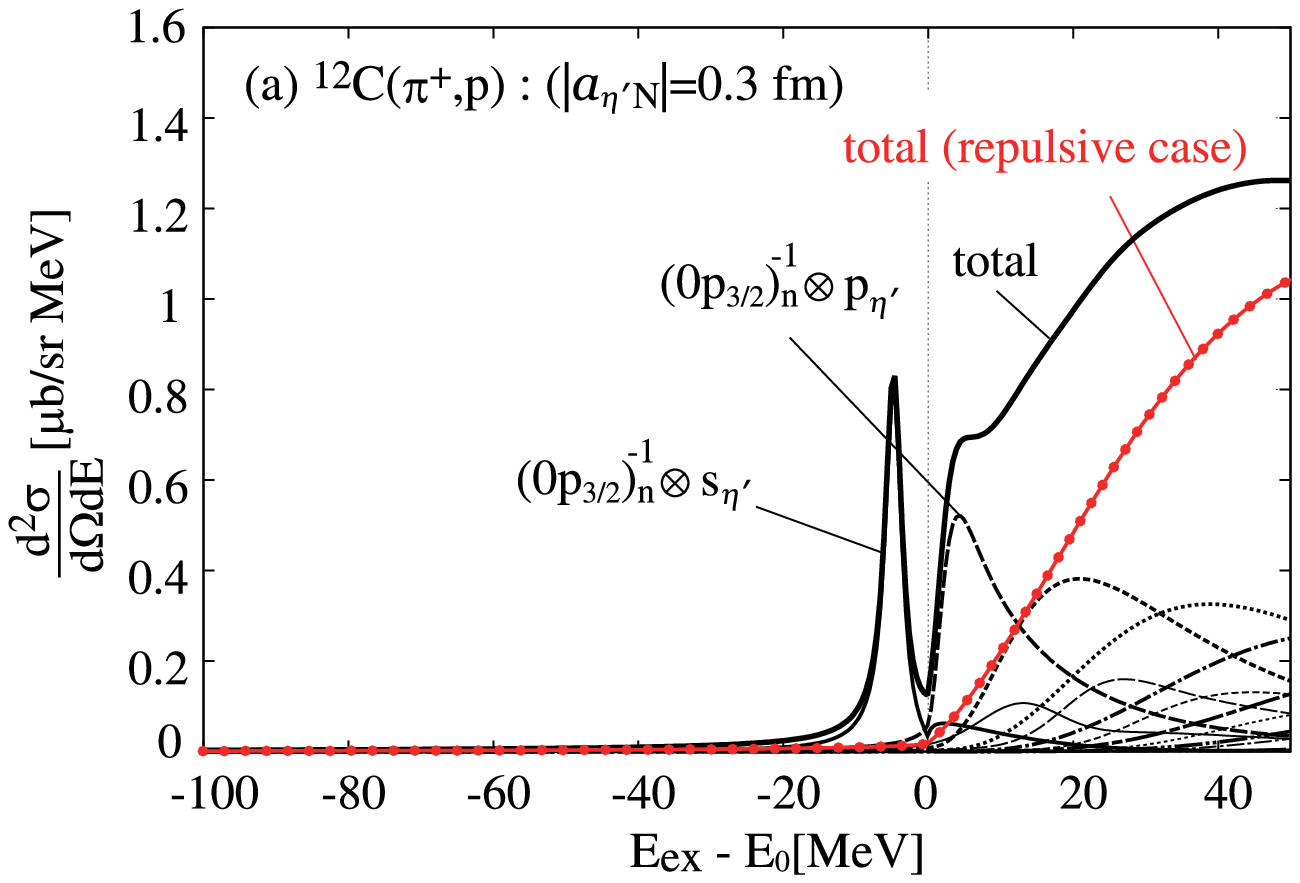}}
\caption{(left) $^3He $ kinetic energy spectrum (not acceptance corrected) in the $^{122}Sn(d,^3He)$ reaction measured with the BigRIPS spectrometer. (right) Theoretically predicted differential cross section for the population of an $\eta^\prime$ mesic state in the $^{12}C(\pi^+,p)$ reaction \cite{Nagahiro_OR}.}
\label{fig:mesic}
\end{center}
\end{figure}
\noindent
Deeply bound $\pi^-$-nucleus states have been identified at the fragment separator FRS at GSI in Pb \cite{Geissel} and Sn isotopes \cite{Suzuki}, populated in the $(d,^3He)$  reaction in almost recoil free kinematics. These states reside in a potential well arising from the interplay of an attractive $\pi^-$ - nucleus Coulomb interaction and the strong interaction which is repulsive for low $\pi^-$ momenta. The 1s binding energies and widths have been determined and used to extract the s-wave pion-nucleus potential. The observed enhancement of the isovector parameter $\mid b_1\mid$ over the free $\pi N$ value has been interpreted as arising from a reduction of the chiral order parameter f$_{\pi}$, thereby providing first experimental evidence for a partial restoration of chiral symmetry in a nuclear medium \cite{Suzuki}. Allowing for a larger range in momentum transfer to the $\pi^-$-nucleus system, also higher lying pionic states have recently been identified in the $\pi^-  - ^{121}$Sn system in experiments at the BigRIPS spectrometer (RIKEN), as shown in Fig.~\ref{fig:mesic} (left) \cite{Itahashi}. These measurements demonstrate the feasibility for detailed systematic studies of the pion-nucleus interaction. For sufficient meson-nucleus attraction due to the strong interaction bound states of neutral mesons and nuclei have been predicted to exist \cite{Liu_Haider}. A search for $\omega$ mesic states is ongoing at CBELSA/TAPS which, however, suffers from the rather strong in-medium broadening of the $\omega$ meson. The relatively narrow in-medium width of about 20 MeV of the $\eta^\prime$ meson \cite{Mariana_etaprime} makes it a promising candidate for a mesic state only bound by the strong interaction. Experiments to search for $\eta^\prime$ mesic states are planned at CBELSA/TAPS \cite{Mariana_proc}, FRS (GSI) \cite{Fujioka_proc}, and JPARC. A theoretically expected spectrum for the population of an $\eta^\prime$ mesic state in the $^{12}C(\pi^+,p)$ reaction is shown in Fig.~\ref{fig:mesic} (right) \cite{Nagahiro_OR}.
  
\section{Conclusion}
\label{sec:6}
In-medium properties of $\omega, \phi$ and $\eta^\prime$ mesons have been studied at several accelerator laboratories. Measured transparency ratios indicate an in-medium broadening to 130-150, 30-50 and 15-25 MeV respectively, extrapolated to normal nuclear matter density.  Inelastic in-medium meson-nucleon cross sections of $\approx$ 60, 15-25, and 3-10 mb, respectively, have been extracted. The analysis of meson line shapes has not provided evidence for an in-medium mass shift.  Several $\pi^-  - ^{121}$Sn bound states have been identified. Encouraged by the relatively narrow $\eta^\prime$ in-medium width of $\approx$ 20 MeV, several experiments are planned to search also for $\eta^\prime$-mesic states. These measurements appear to be a promising tool for systematic studies of the meson-nucleus interactions and in-medium effects.

\end{document}